# Quantification of cell contractile behavior based on non-destructive macroscopic measurement of tension forces on bioprinted hydrogel

**Authors**

Sarah Pragnere[1]
Naima El Kholti[2]
Leslie Gudimard[3]
Lucie Essayan[3]
Christophe Marquette[3]
Emma Petiot[3]
Cyril Pailler-Mattei[1,4]

Addresses:
[1] Laboratory of Tribology and System Dynamics UMR-CNRS 5513, Ecole Centrale de Lyon, France

[2] UMR 5305, Tissue Biology and Therapeutic Engineering Laboratory (LBTI), University of Lyon, CNRS, 69367 Lyon, France.

[3] 3d.FAB, Univ Lyon, Université Lyon1, CNRS, INSA, CPE-Lyon, ICBMS, UMR 5246, 43, Bd du 11 Villeurbanne cedex, France

[4] University of Lyon, Université Claude Bernard Lyon 1, ISPB-Faculté de Pharmacie de Lyon, France

Corresponding authors :
Sarah Pragnere: sarah.pragnere@ec-lyon.fr
Cyril Pailler-Mattei: cyril.pailler-mattei@ec-lyon.fr

**ABSTRACT**

Contraction assay based on surface measurement have been widely used to evaluate cell contractility in 3D models. This method is straightforward and requires no specific equipment, but it does not provide quantitative data about contraction forces generated by cells. We expanded this method with a new biomechanical model, based on the work-energy theorem, to provide non-destructive longitudinal monitoring of contraction forces generated by cells in 3D.

We applied this method on hydrogels seeded with either fibroblasts or osteoblasts. Hydrogel mechanical characteristics were modulated to enhance (condition $HCA_{High}$: hydrogel contraction assay high contraction) or limit (condition $HCA_{Low}$: hydrogel contraction assay low contraction) cell contractile behaviors. Macroscopic measures were further correlated with cell contractile behavior and descriptive analysis of their physiology in response to different mechanical environments. Fibroblasts and osteoblasts contracted their matrix up to 47% and 77% respectively. Contraction stress peaked at day 5 with $1.1 \cdot 10^{-14}$ Pa for fibroblasts and $3.5 \cdot 10^{-14}$ Pa for osteoblasts, which correlated with cell attachment and spreading. Negligible contraction was seen in $HCA_{Low}$. Both fibroblasts and osteoblasts expressed α-SMA contractile fibers in $HCA_{High}$ and $HCA_{Low}$. Failure to contract $HCA_{Low}$ was attributed to increased cross-linking and resistance to proteolytic degradation of the hydrogel.

**ABBREVIATIONS**

α-SMA: α-Smooth Muscle Actin

$HCA_{high}$: Hydrogel Contraction Assay high contraction

$HCA_{low}$: Hydrogel Contraction Assay high contraction

ECM: Extra-cellular Matrix

FBS: Foetal bovine Serum

MMP: matrix metalloproteases

**INTRODUCTION**

3D cell culture has been used for several decades to develop *in vitro* biological models and to study cell behavior for applications in pharmacology and regenerative medicine. This is notably useful to treat burns or chronic wounds with skin substitutes [1–4], or to repair critical-sized bone defects [5]. 3D cell culture can also provide models to study of a variety of biological functions and behaviors like the responses to mechanical or biochemical stimuli [6–9], to cellular ageing [10] or the impact of genetic disease [11].

Two different approaches co-exist for 3D cell culture: cell seeding on porous scaffolds [3,10,12,13] or cell embedding in hydrogels made of biocompatible hydrogels [8,14,15]. Among these hydrogels, collagen has been extensively studied, being one of the most abundant proteins of the extracellular matrix (ECM) in connective tissues [16], including dermis [17], bone [18], cartilage [19] and tendon [20].

ECM impacts cellular fate by providing mechanical support and attachment for cells [16]. *In vivo*, it is secreted by tissue resident cells [16], and is constantly remodeled. ECM is successively degraded by matrix metalloproteases (MMP) and neo-synthetized [21]. 3D cell culture models aim at reproducing such mechanisms to replace existing artificial scaffolds or hydrogels with an *in-vivo* like matrix.

During cultivation of 3D *in vitro* models, cells generate tractional forces. When they are seeded on compliant materials such as collagen hydrogels, such forces reshape the structures surrounding the cells [22,23]. This results in changes of hydrogels volume, mostly along the radial direction [24], which is known as matrix contraction [22]. Several resident cells can contract their matrix such as dermal fibroblasts [25], osteoblasts [26], chondrocytes [27] and tenocytes [28].

Matrix contraction involves two main mechanisms, which both require cells to possess an actin cytoskeleton [22,29]. First cellular behavior is cell traction on matrix fibers during cell attachment and locomotion. This happens during the first phase of wound healing when fibroblasts migrate towards the damaged tissues [30]. This has also been observed *in vitro* for fibroblasts embedded in free-floating collagen hydrogels [31]. The second cellular behavior is cell contraction (*i.e.* cell shortening) transmitted to the tissue. It requires expression of extracellular α-Smooth Muscle Actin (α-SMA) fibers and strong focal adhesions [32–34].

Although studies focused mainly on fibroblasts, osteoblasts in collagen hydrogel showed greater contractability [35]. *In vitro*, expression of α-SMA by fibroblasts was observed on deformable silicon membrane, generating significantly higher material deformation than non-expressing cells [34]. Osteoblasts were also described to transiently express α-SMA during fracture healing [36] and to contract collagen hydrogels sufficiently to bring in contact two pieces of scaffold embedded in the same collagen hydrogel [5].

Contraction of collagen alters the structure of a tissue and the behavior of the resident cells. As connective tissues provide support for other tissues, matrix contraction can impact body physiology [22]. Contractility is notably involved in wound healing [22,32] and in fracture healing [36]. Dysregulation of matrix contraction can lead to hypertrophic scar [30] or cancer [24,37]. Thus, it is of major interest to understand how and why cells interact with their environment.

Collagen hydrogel contraction *in vitro* was first mentioned by Elsdale and Bard in 1972. They noticed a "gradual collapse of the lattice to a dense, opaque body less than one-tenth of the original size" of their collagen hydrogel seeded with fibroblasts [38]. Since then, it has been studied at macroscopic and microscopic scales. Bell and co-workers introduced collagen contraction assays to evaluate fibroblasts behavior in collagen hydrogels [25]. This macroscopic method has been widely applied as it requires only images of hydrogels or scaffolds to monitor surface area [4,23,24,26,29,39,40]. It has the benefit to be non-destructive and supports longitudinal monitoring. However, this method provides merely qualitative data with percentage of contraction but no quantitative data on the contraction forces generated by cells. To overcome such limitation, a first set of methods focused on deformation of 2D supports by the cells, such as deformable membrane [41], traction force microscopy [42] or elastic micro-pillar [43]. Although they provide quantitative data on forces generated by individual cells, these techniques are limited to 2D cultures and failed to be transferred to 3D environment to mimic *in vivo* traction forces. Traction force microscopy has been adapted to 3D cell culture system [44] but it is mostly used with purely elastic materials that do not reproduce the biochemical and mechanical properties of ECM [23,24]. Another major drawback is the inherent evolutive behavior of cells. During the remodeling process, cells are concomitantly degrading their support and synthesizing their own ECM [44], which induce strong changes in material support mechanical properties [45,46]. As precise knowledge about material elasticity and viscosity is required, traction force microscopy has been limited to 90 minutes studies [44,47].

Culture force monitoring devices have been developed to directly measure macroscopic forces generated by cells in 3D [48–51]. Hydrogels are clamped to strain gauges that measure the contraction forces generated by cells. This technique provides useful quantitative follow-up but the system setup applies a static tensional force on cells, which is known to influence cell behavior [21,52,53]. This model reproduces wound healing whereas free-floating hydrogels are closer to dermis model [22].

Consequently, after 40 years of methods development, there is still a need for a non-destructive technique to monitor quantitatively cell contraction during *in vitro* cultivation of 3D models which could closely imitate *in vivo* tissue behaviors.

We aimed to develop a new biomechanical model to non-destructively quantify cell contraction forces. Hydrogels seeded with fibroblasts or osteoblasts, known for their contractile behavior, were monitored by longitudinal analysis of hydrogel surface over 35 days. Cylindrical hydrogels were 3D bioprinted to procure an isotropic biomaterial. Hydrogels supporting cell growth were modulated in term of mechanical characteristic to enhance or limit cell contractile behaviors. Macroscopic measures were further correlated with cellular contractile behavior and descriptive analysis of their physiology in response to different mechanical environments.

**MATERIAL AND METHODS**

- Cells' isolation and cultivation

Foreskin fibroblasts [15] were cultivated in flasks in Dulbecco's modified Eagle medium (DMEM)/Glutamax medium (Gibco$^{TM}$ #31966021), supplemented with 10% foetal bovine serum (Gibco$^{TM}$ # A3160802). Primary human osteoblasts were purchased from PromoCell (#C-12720) and grown in flasks with DMEM low glucose (Dutscher #L0064) supplemented with 10% fœtal bovine serum (PAN-Biotech #500105) and 50µg/mL ascorbic acid (Sigma #A8960). All 2D cultured cells were grown in an incubator at 37°C and 5% $CO_2$.

- Bio-Ink Formulation & bioprinting protocols

The isotropic shapes of tissue constructs were designed as cylinder geometries of 10mm diameter and 1.6mm thickness. Based on an in-house patented formulation [15,54,55], the bio-ink was formulated as a mixture of 5% (w/v) bovine gelatin (Sigma #G1890), 2% (w/v) very low viscosity alginate (Alpha Aesar # A18565), and 2% (w/v) fibrinogen (Sigma # F8630) dissolved overnight in DMEM without calcium (Gibco™ #21068028) at 37 °C. Just before printing, fibroblasts or osteoblasts were trypsinized and suspended in the bio-ink to obtain $3 \times 10^5$ cells mL$^{-1}$. After homogenization, a 10ml sterile syringe (Nordson EFD) was loaded and incubated at room temperature. Tissue constructs were consolidated by 60 min incubation in either a solution $HCA_{high}$ containing 3% w/v CaCl2, 0.2% w/v transglutaminase (Ajinomoto ActivaWM) and 20 U/mL of thrombin (Sigma # T4648-10KU) or a solution $HCA_{low}$ containing DMEM medium with 0.02% calcium, 4% w/v transglutaminase and 20 U/mL thrombin. Following consolidation step, constructs were then rinsed three times with sterile NaCl 0.9% (Versol).

- Tissue cultivation & monitoring

After the consolidation step, tissue constructs were grown in 12-well plates for 35 days at 37°C in a 5% $CO_2$ atmosphere. Culture medium was renewed twice a week. For the dermis model, freshly prepared ascorbic acid at 50µg/mL was added at each medium renewal. For the bone constructs, osteogenic medium consisted in 50µg/mL ascorbic acid, 5mM β-glycerophosphate (Sigma #G9422) and 100 nM dexamethasone (Sigma # D4902).

Tissue construct contraction was monitored by macroscopic images of the tissues. ImageJ was then used to determine the tissue construct area in reference of well area of the culture plates. Well and tissue construct areas were measured in pixels by manual contouring. The known area in square centimeters of the well was used to set scale and convert square pixels in square cm (Supplementary S1). At least 3 tissue constructs were measured at every time point.

Cell viability and proliferation were assessed by calcein-AM staining (Invitrogen™ #C1430). Tissue constructs were incubated with 0.5mL of 2μM calcein-AM for 30 minutes at 37°C, washed in PBS and imaged using a fluorescence microscope (Nikon Eclipse Ts2R) equipped with a 475nm filter. To evaluate proliferation, cell growth kinetics were obtained by cell counting after tissue constructs enzymatic dissolution. Tissue constructs were washed in PBS and incubated 10 minutes in 1mL of 0.05% trypsin. 1mL medium supplemented with FBS was used to stop the trypsin action. After pelleting by centrifugation, tissue constructs were washed with 2mL PBS and further incubated in 30mg/mL collagenase A (Merck # COLLA-RO). Vigorous shaking was applied every five minutes until complete dissolution of the tissue construct, which commonly occurs within 10 to 20 minutes. After a last pelleting by centrifugation, extracted cells were resuspended and counted using a hemacytometer.

- Characterization of cell morphology

Tissues were fixed overnight in paraformaldehyde at 4°C. They were then dehydrated in a gradient of ethanol, followed by paraffin embedding and cut into 5μm sections. Phalloidin staining of actin filaments in the tissues with fibroblasts was performed on histological sections after deparaffinization and rehydration. Samples were permeabilized in acetone at -20°C for 3 minutes then washed two times in deionized water. Sections were incubated with Alexa Fluor™ 546 Phalloidin at 5 U/mL for 45 minutes to stain the actin cytoskeleton and nuclei were counterstained with 300nM DAPI dilactate for 5 minutes. Images were acquired with an Axio Scan.Z1 (Zeiss) equipped with a 20X objective and 405 nm and 561 nm filters. In case of osteoblasts, analysis was performed by confocal microscopy right after paraformaldehyde fixation. Tissues were permeabilized with 0.1% TritonX-100 for 15 minutes, washed in PBS then incubated with Alexa Fluor™ 546 Phalloidin (Invitrogen™ #A22283) at 5 U/mL for 90 minutes to stain the actin cytoskeleton and 5 minutes with 300nM DAPI dilactate (Invitrogen™ #D3571) to stain the nuclei. This allowed us to better visualize few numbers of cells which were ten times lower than fibroblasts. Confocal scans were acquired using a Zeiss LSM 880 microscope equipped with 40X immersion objective and 405 nm and 561 nm filters. Maximum intensity images were generated from z-stacks with a distance of 2μm between each slice.

- Characterization of cell phenotypes

After deparaffinization and rehydration, antigen retrieval was performed in 10mM citrate buffer (pH 6) for 20 minutes at 100°C. The following steps were performed with the ImmPRESS® Excel Amplified Polymer Kit (Vector Laboratories, MP-7602) according to manufacturer's instructions. Endogenous peroxidases were blocked with BLOXALL for 10 minutes, and non-specific binding sites were blocked by incubation with BlockAid blocking solution (ThermoFisher, B10710) for 30 minutes. Primary antibody (α-SMA 1:800, Sigma, A2547) was incubated overnight at 4°C. After washing in PBS, Amplifier Antibody (goat anti mouse) was incubated for 15 minutes followed by 30 minutes incubation with ImmPRESS Excel Amplified HRP Polymer Reagent. Revelation was performed using ImmPACT® DAB *EqV* Substrate. Nuclei were counterstained with Gill's Hematoxylin.

- Mechanical properties analysis by Micro-indentation

After one night in the incubator to equilibrate, mechanical properties of the tissue constructs were measured by an in-house developed light load indentation device [45]. Indentation tests were carried out at constant applied normal load $F_z$=1 mN and constant indentation speed $V$=25 µm.s$^{-1}$. The indenter used was a spherical PTFE indenter, with a radius of curvature $R$= 1.6 mm. Measurements were repeated at least 3 times per sample and each type of sample was indented in duplicate.

**RESULTS & DISCUSSION**

Since Bell and co-workers introduced collagen gel contraction assay [25], it has been extensively used to study fibroblasts [4,9,56,57] or osteoblasts [5,29,40] ability to contract a collagen hydrogel. We applied a similar methodology on hydrogels made of gelatin, alginate and fibrinogen containing either fibroblasts or osteoblasts. We used hydrogel contraction assay (HCA) to monitor tissue constructs macroscopic contraction and we correlated it to morphology of actin cytoskeleton and expression of α-SMA fibers. To identify different cell biological behaviors, we selected two hydrogel conditions: one allowing a high degree of contraction, named HCA$_{High}$, and one limiting the contraction effect, named HCA$_{Low}$.

    I. **Cross-linking impact on mechanical properties of synthetic ECM / biomatrix**
        1. **Mechanical properties of hydrogels**

Hydrogel formulation and consolidation protocols were adapted from Pourchet et al. [15] as they were known to provide good fibroblast proliferation. Mechanical properties were hardly measured with the original formulation and consolidation conditions as tissue constructs collapsed during indentation measurement. A new reticulation protocol was tested with the addition of 4% transglutaminase enzyme to reticulate gelatin of the hydrogel formulation. Hydrogel elastic modulus thus reached 32.0 10$^3$ Pa ± 99 Pa allowing for indentation measurement. Nevertheless, no cell proliferation and tissue construct contraction were observed. These preliminary assays demonstrated that hydrogel biomechanical properties must be in close range allowing both cell proliferation and sample handling to measure contraction forces. On this basis, the two following combinations were satisfactory to achieve an HCA assay on fibroblast and osteoblast tissue constructs: the HCA$_{low}$ condition, consisting in a hydrogel consolidated with 4% transglutaminase and 0.02% CaCl2 contained in DMEM and the HCA$_{high}$ condition, consisting in hydrogel consolidated with 0.2% transglutaminase and 3% CaCl2. Both presented sufficient stiffness to obtain mechanical characterization and cell proliferation. Elastic modulus of HCA$_{High}$ were 6.3 10$^3$ Pa ± 1.3 10$^3$ Pa for fibroblasts and 3.7 10$^3$ Pa ± 755 Pa for osteoblasts. Elastic modulus of HCA$_{Low}$ were in the same range with 7.0 10$^3$ Pa ± 1.4 10$^3$ Pa for fibroblasts and 6.3 10$^3$ Pa ± 192 Pa for osteoblasts. These values were in the same range as those observed for skin [58]. The newly synthesized bone matrix, produced in vitro by an osteoblast cell line, displayed an elastic modulus of 27 kPa ± 10 kPa [59]. Our constructs with similar stiffness did not allow for a complete cell differentiation, which is why we chose to work with milder polymerization. Compared to collagen hydrogels that have elastic modulus from 2 Pa to 720 Pa [35,50,60,61] our constructs are still closer to the natural bone matrix before mineralization occurs, which make them valuable for further biomechanical studies of bone cells (figure 1).

        2. **Hydrogel contraction**

Contraction of hydrogels was monitored over 35 days by macroscopic images (figure 2 A and E). This method has been extensively used to study cell behavior in 3D hydrogels but did not provide quantitative data about traction force generated by the cells [9,25,29,56]. Therefore, we developed a biomechanical model based on the kinetic energy theorem to quantify contraction forces from measurements of tissue construct surfaces.

Small variations in the surface of samples without cells were attributed to measurement variations and interactions between the hydrogel and the culture medium. Acellular $HCA_{High}$ constructs degraded and collapsed within the 12 first days of incubation in either osteogenic or fibroblastic medium. We thus used the respective acellular hydrogel to normalize the tissue construct longitudinal contraction. Hydrogel contraction by both osteoblasts and fibroblasts depended on the polymerization condition. After 35 days of culture, fibroblasts and osteoblasts had contracted their hydrogel to 47% and 77% (supplementary figure S2) respectively of their initial surface in $HCA_{High}$ while in $HCA_{Low}$, the contraction was 24% for fibroblasts and 14% for osteoblasts.

To quantify the traction force generated by the cells responsible for the contraction of the hydrogels, we applied the work-energy theorem to the evolution of tissue construct surfaces.

The work-energy theorem is defined as:

$$W_{\vec{F}_{ext}} + W_{\vec{F}_{int}} = \Delta E_K \qquad (1)$$

where $W_{\vec{F}_{ext}}$ is the work of the external forces, $W_{\vec{F}_{int}}$ is the work of the internal forces and $\Delta E_K$ is the studied system change of kinetic energy, namely the tissue construct. We considered that cells were mainly responsible for the contraction of the tissue construct. Consequently, the work of external forces was neglected compared to the work of internal forces: $W_{\vec{F}_{ext}} \ll W_{\vec{F}_{int}}$

Hence, we approximated the change of kinetic energy by:

$$W_{\vec{F}_{int}} = \Delta E_K \qquad (2)$$

Cells produced on a small surface $ds$ on the sidewall of the tissue a small force $d\vec{F}_{int}$ such as: $\vec{F}_{int} = -\tilde{p}ds\vec{e}_r$, where $\tilde{p}$ is the stress applied by the cells onto the surface $ds$ and $\vec{e}_r$ is the radial vector in cylindrical coordinates (figure 3). In the following, we assumed that cells acted mainly on the sidewall of the tissue construct, meaning that we supposed that the height of the tissue samples $h$ remained constant [62].

The work done by the internal forces $W_{\vec{F}_{int}}$ when the system moves an infinitesimal amount $dr$ is defined as:

$$W_{\vec{F}_{int}} = \int d\vec{F}_{int} \cdot dr\vec{e}_r = \int -\tilde{p}ds\vec{e}_r \cdot dr\vec{e}_r = \int -\tilde{p}dsdr$$

In cylindrical coordinates, the small surface $ds$ is expressed as: $ds = rd\theta dz$, so the work done by the internal forces is given by:

$$W_{\vec{F}_{int}} = \int -\tilde{p}\, r\, dr\, d\theta\, dz$$

We assumed that the stress applied by cells on the sidewall of the tissue was homogenous, isotropic and constant for the given period of time, hence:

$$W_{\vec{F}_{int}} = -\tilde{p}\int_{r_0}^{r_i} r\,dr \int_0^{2\pi} d\theta \int_{-\frac{h}{2}}^{+\frac{h}{2}} dz = -\pi \tilde{p} h (r_i^2 - r_0^2) \qquad (3)$$

Where $r_i$ is the radius of the sample at time $i$ and $r_0$ the radius of the sample at initial time.

In classical mechanics, the change of kinetic energy for a point object of mass $m$ is given by:

$$\Delta E_K = \frac{1}{2} m v_i^2 - \frac{1}{2} m v_0^2 \qquad (4)$$

where $v_i^2$ and $v_0^2$ are respectively the square of the speed at time $i$ and initial time. In our study, a variation of the volume of biological sample has been observed as a function of time. We assumed that the radius of the sample was much higher than the height ($h \ll r$), hence, by analogy to the change of kinetic energy for a point object, we write the change of kinetic energy for our sample as:

$$\Delta E_K = \frac{1}{2} \sigma \left( \overline{s_i}^2 - \overline{s_0}^2 \right) \qquad (5)$$

where $\sigma$ is the surface mass of the sample and $\overline{s_i}^2$ and $\overline{s_0}^2$ are respectively the square of the surface speed at time $i$ and initial time. In our study, the initial conditions involved that the surface speed at the initial time was equal to 0, so eq. 5, is rewritten as:

$$\Delta E_K = \frac{1}{2} \sigma \overline{s_i}^2 \qquad (6)$$

The surface speed at time $i$ is defined as: $\overline{s_i} = \frac{s_i - s_0}{t_i - t_0}$, where $s_i$ and $s_0$ are respectively the surface of the biological sample at time $i$, noted $t_i$, and initial time, noted $t_0$.

The volume of the sample at $t_i$, is defined as: $v_i = s_i h$, and the volume of the sample at initial time $t_0$ is: $v_0 = s_0 h$.

As we assumed that the contraction of the tissue happened in the radial direction and that the height $h$ of the tissue remained constant, the surface speed at time $i$, is then expressed as: $\overline{s_i} = \frac{1}{h}\left(\frac{v_i - v_0}{t_i - t_0}\right)$. The change of kinetic energy (Eq.6) is defined as:

$$\Delta E_K = \frac{1}{2} \frac{\sigma}{h^2} \left( \frac{v_i - v_0}{t_i - t_0} \right)^2 \qquad (7)$$

The ratio $\frac{v_i - v_0}{t_i - t_0}$ is homogenous to the volume speed of the sample at time $i$, noted $\overline{v_i}$. Moreover, the ratio $\frac{\sigma}{h}$ is equivalent to the density of the sample, noted $\rho$. As a consequence, Eq.7 can be rewritten as:

$$\Delta E_K = \frac{1}{2}\frac{\rho}{h}\overline{v}_i^2 \tag{8}$$

From Eq.2 using equations (3) and (8), the stress $\tilde{p}$ applied by the cells in the tissue constructs is given by:

$$\tilde{p} = -\frac{\rho}{2\pi h^2}\frac{\overline{v}_i^2}{(r_i^2-r_0^2)} = -\frac{\rho}{2}\frac{s_i-s_0}{(t_i-t_0)^2} \tag{9}$$

Cell contraction was determined after normalization by traction stress ($\tilde{p}$) evaluated on the respective acellular control hydrogels. $HCA_{Low}$ did not demonstrate a significative cell traction stress over the 35 days of culture for both cell types (Figure 2 B and F). Oppositely, a positive traction stress was observed for the $HCA_{High}$ constructs. Maximal values were reached at day 5 with $1.1 \cdot 10^{-14}$ Pa for fibroblasts and $3.5 \cdot 10^{-14}$ Pa for osteoblasts. Then, traction stress decreased over time, dropping to $3.0 \cdot 10^{-15}$ Pa for fibroblasts and to $4.0 \cdot 10^{-15}$ Pa for osteoblasts after 35 days of culture. Osteoblasts showed a temporary increase of traction stress between day 12 and 20, corresponding to the elevation of the rate of surface contraction.

To correlate traction stress to biological behavior, we first monitored the cell growth kinetics of both fibroblasts and osteoblasts. This also allowed us to calculate the specific traction, *i.e.* traction generated per cells, in both condition over time. Growth kinetics were performed thanks to tissue construct dissociation at days 12, 22 and 35.

The global increase of contraction stress was not only a consequence of an increased number of cells in the tissue construct but also an increased contractility of cells. Indeed, the specific traction generated per cell was higher for both fibroblasts and osteoblasts in $HCA_{High}$ biomaterial than in $HCA_{Low}$ at all time points. This difference reached $2.2 \cdot 10^{-20}$ Pa for fibroblasts at day 12 and $3.0 \cdot 10^{-19}$ Pa for osteoblasts at day 22. Interestingly, specific traction is different in magnitude and evolves differently depending on the cell origin. Osteoblasts in $HCA_{High}$ generated 11 to 25 times higher traction stress than fibroblasts in the same condition. Specific traction of fibroblasts decreased steadily of 75% between day 12 and 35 while on the contrary osteoblasts intensified their specific traction between day 12 and 22 to later decrease again between day 22 and day 35. Past studies have already shown differences in contractile ability between cell types and species [49,63,64] and even between fibroblasts from dermis or foreskin [50].

In studies using collagen gels, most of contraction happens during the first 24h of culture then reach a plateau [25,29,39,51,57,64]. Forces generated by single cells range from 0.2nN [49,64] to 25nN [62]. This corresponds to attachment of cells to the matrix and formation of cell cytoplasmic extension [49]. We computed specific contraction forces generated per cells by normalization of the traction stress with the surface area. In our case, the maximum forces generated at day 5 were 0.2nN for fibroblasts and 0.7nN for osteoblasts. This fell in the lower range of values found in literature [25,65], which correspond to hydrogel with elastic modulus of 0.1 kPa. In our case, hydrogel elastic modulus was 6.2 kPa for fibroblasts and 3.7 kPa for osteoblasts, which was 37 to 62 times higher. As it has been shown that matrix stiffness (here defined as opposite of compliance) lowers contraction by cells [25,65], this explains the lower specific force values that we obtained.

Differences between the two tissue constructs $HCA_{High}$ and $HCA_{Low}$ in term of contraction forces confirm the hypothesis that hydrogel resistance is strongly impacting contraction forces and cell behavior. Since mechanical properties at day 1 were similar between $HCA_{High}$ and $HCA_{Low}$, the difference of measured contraction must come from the ability of the cells to reorganize the matrix. Polymerization with transglutaminase cross-links gelatin through ε(γ-glutamyl)lysine linkage using the side chains of lysine and glutamine residues. This enhances hydrogel resistance to proteolytic degradation [66]. Is thus likely that cells cannot remodel their surrounding matrix in hydrogels polymerized with high levels of transglutaminase. This is further supported by the fact that acellular $HCA_{High}$ samples degraded during cultivation whereas acellular $HCA_{Low}$ samples remained intact during the 35 days of cultivation.

## II. Hydrogel biomechanical impact on cell physiological behavior

Macroscopic differences observed on generated contraction forces were linked with the characterization of cell growth profile in the matrix and with their contractile phenotypes as for example the expression of actin [22,29] and specialized contractile α-SM actin [34,67].

### 1. Cell viability and spreading

For the first time in a contraction study, the growth pattern of fibroblasts and osteoblasts were monitored by calcein-green live cell staining and cell counting after tissue constructs dissociation (figure 2 C and G and figure 4) to follow cell proliferation trends and cell morphology within the hydrogels.

Concerning growth kinetics, both cell types displayed increased growth rate in $HCA_{High}$ compared to $HCA_{Low}$. Fibroblasts showed similar growth trends for both constructs with a linear increase until day 21, corresponding to a growth rate of 0.17 $day^{-1}$ in $HCA_{High}$ and 0.13 $day^{-1}$ in $HCA_{Low}$. This growth was followed by a stabilization phase at a cell density 2.4 times higher in $HCA_{High}$ than in $HCA_{Low}$. Cell proliferation was more limited for osteoblasts with a growth rate of 0.07 $day^{-1}$ in $HCA_{High}$ and no growth in $HCA_{low}$ conditions. Again, cell densities in $HCA_{High}$ were 2 times higher than in $HCA_{Low}$ after 35 days of culture.

To detail growth behaviors of both cells in these hydrogels, cells morphology and repartition were followed by live cells calcein staining. Three growth phases were identified for fibroblasts. First, within the 5 first days, fibroblasts displayed a spherical shape. From day 12 they started to stretch and demonstrated stellate shapes. An interconnected network developed and it was only after 21 days of culture that fibroblasts formed a dense layer on the construct periphery which remained until the end of the culture. Additional staining of cell nuclei with DAPI and of actin cytoskeleton with phalloidin on histological cuts revealed differences in cell repartition inside the construct depending on the polymerization condition (figure 5 and supplementary S3). Fibroblasts in $HCA_{High}$ were well distributed throughout the hydrogel and were connected to each other through their cytoplasmic extensions while in $HCA_{Low}$ cells were mainly distributed on the surface of the construct with a polarized morphology. Very few cells exhibiting cytoplasmic extensions were distributed inside the construct. In both HCA$_{High}$ and HCA$_{Low}$ cells displayed the fibroblasts morphology described in literature for free floating hydrogels, thus confirming the phenotype of contractile cells [33].

Three growth phases are also visible for osteoblasts. From day 1 to day 5, osteoblasts displayed a cuboidal morphology characteristic of active cells that are not fully differentiated. From day 12, few cells initiated some dendrites. This proved the beginning of differentiation process into osteocytes. After 35 days of culture, cells formed an interconnected network through their dendrites, demonstrating a fully differentiated state into osteocytes. This was further supported by expression of osteocytic marker PHEX (supplementary S4). DAPI and phalloidin staining confirmed these assumptions. Cells exhibited a cuboidal morphology with some osteoblasts starting to extend cytoplasmic prolongations for both conditions at day 12 (figure 4). However, after 33 days of culture, strong differences between constructs were observed. HCA$_{High}$ demonstrated a higher internal cell density with osteocytes interconnected thanks to their dendrites. On the contrary, in the HCA$_{Low}$ tissue constructs, cells were mostly distributed on the periphery and showed an elongated morphology with some prolongations. Besides, cells did not appear to form a dense interconnected network like for osteocytes present in the HCA$_{High}$ constructs.

Both osteoblasts and fibroblasts reached a maximum of global traction stress at day 5. This correlated with attachment of cells (see figure 2 day 5). This is also relevant with maximal contraction in collagen gels during the attachment phase of the cells [49,63].

2. **Expression of α-SMA**

Cell contractile phenotype is identified thanks to the detection of α-SMA fibers. Thus, expression of α-SMA fibers was compared for both cells in HCA$_{High}$ and HCA$_{Low}$ conditions. Both fibroblasts and osteoblasts in HCA$_{High}$ condition expressed α-SMA associated with a strong contraction of the biomatrix (47% and 77% respectively). A clear biomatrix remodeling was also observed when compared with HCA$_{Low}$ and was further confirmed by collagen staining (supplementary S5). Interestingly, both osteoblasts and fibroblasts also expressed α-SMA in HCA$_{Low}$. This proved that cells in HCA$_{High}$ and HCA$_{Low}$ presented a contractile behavior. However, as cells reached a lower density in HCA$_{Low}$ than in HCA$_{High}$ and were distributed mainly on the construct perimeter, contraction occurred only on the hydrogel periphery. In contrast, cells in HCA$_{High}$ applied contraction forces homogeneously on the overall construct volume.

Thanks to labelling of α-SMA fibers, we confirmed fibroblasts transition to myofibroblasts described in wound contraction. This behavior is generally initiated by surrounding fibroblasts migration to the wound and is described as a mechanism which stiffened the ECM [33].

As for osteoblasts seeded hydrogels, we identified several transition phases for cell phenotypes through expression of α-SMA fibers. Expression of α-SMA by osteoblasts has already been demonstrated both *in vitro* and *in vivo* [26,36,68]. Active osteoblastic cells involved in fracture repair express α-SMA fibers while it is no longer the case for mature osteocytes. This is consistent with the observed behavior in our experiment with a proliferation phase occurring during the first 12 days when cells exhibited expression α-SMA fibers and generated high cell specific contraction (figure 2 and 6), while the second growth phase until 35 days of culture demonstrated an osteocyte phenotype and a stabilization of contraction forces.

An advantage of our study compared to the ones presented in literature is the monitoring of contraction forces over long cultivation periods - 35 days in the present case. This allowed us to clearly identify different cellular processes involved in biomatrix contraction. Cell growth was only characterized over the first 21 days of culture for fibroblasts and 12 days for osteoblasts. This meant that novel cell attachment could not account for the contraction occurring after the growth phases. Expression of $\alpha$-SMA suggested a shift of contraction mode from cell attachment and locomotion to cell shortening. Cells in $HCA_{High}$ generated specific contraction stress an order of magnitude higher than in $HCA_{Low}$ but no differences were noticed in contractile phenotype of the cells. This indicated that hydrogel resisted to cell contraction forces. As contraction depends on cell density [25,35], cells in $HCA_{High}$ probably did not reach the density threshold to initiate contraction. To verify such hypothesis, further experiments should investigate the effect of a higher seeding cell density on the construct contraction.

III. Conclusion

This study proposes a new method to quantify traction forces generated by cells embedded in hydrogels. This biomechanical model is non-destructive and requires only macroscopic images of the seeded hydrogels and allows for long term longitudinal monitoring of several contractile cell processes. Two contractile cells were studied, fibroblasts and osteoblasts embedded in proliferative hydrogels, over 35 days. Differential reticulation process confers to the selected hydrogels modulable mechanical properties and degradability by cells MMP's. It was possible to enhance cell proliferation and their contractile fibers expression to observe a strong matrix contraction by the cells (up to 50% for fibroblasts and 75% for osteoblasts). On the contrary, the choice of a matrix presenting increased reticulation of the hydrogels, thus impeding cell proliferation, also limited the overall expression of cell contractile fibers and macroscopic observation of such biological function. Future studies should focus on solutions to take into account anisotropy of the traction stress that arises locally inside the hydrogels.

**SUPPLEMENTARY DATA**

**S1: Surface measurement of bioprinted constructs.** Surfaces of the construct (continuous line) and culture well (dashed line) have been measured in pixels. The known area in cm² of the culture well has been used to convert the construct area in cm². $HCA_{High}$: hydrogel contraction assay with high contraction, $HCA_{Low}$: hydrogel contraction assay with low contraction.

**S2: Evolution of hydrogel surface monitored by image analysis expressed as a percentage of surface value at day 1.** $HCA_{High}$: hydrogel contraction assay with high contraction, $HCA_{Low}$: hydrogel contraction assay with low contraction.

**S3: Repartition of fibroblasts at day 33.** Dashed line delineates the perimeter of the tissue construct. Nuclei were stained with DAPI and appear in blue. HCA$_{High}$: hydrogel contraction assay with high contraction, HCA$_{Low}$: hydrogel contraction assay with low contraction.

**S4: Expression of PHEX revealing differentiation from osteoblasts to osteocytes at day 33.** Cells expressing PHEX appear in brown after HRP revelation while cell nuclei are visualized with a counterstaining with Gill's Hematoxylin in dark blue. HCA$_{High}$: hydrogel contraction assay with high contraction, HCA$_{Low}$: hydrogel contraction assay with low contraction.

**S5: Collagen secretion by osteoblasts at day 33.** Collagen appear in brown after immunostaining and revelation by HRP. Osteoblasts in HCA$_{High}$ have remodeled their matrix and secreted a dense collagen matrix while secretion of collagen in HCA$_{Low}$ is localised immediately arround cells. HCA$_{High}$: hydrogel contraction assay with high contraction, HCA$_{Low}$: hydrogel contraction assay with low contraction.